\documentclass[aps,prd,twocolumn,a4paper,showpacs,nofootinbib]{revtex4}
\usepackage{amsmath}
\usepackage{amssymb}
\usepackage{graphicx}
\usepackage{epsfig}
\usepackage{varioref,xr-hyper}
\usepackage[sort&compress]{natbib}
\usepackage{hypernat}
\usepackage[hypertex,draft=false,verbose=false,debug=false]{hyperref}

\newcommand{\EG}{e.g., }
\newcommand{\IE}{i.e., }
\newcommand{\ETC}{etc. }

\newcommand{\ads}{$\mathrm{AdS}_{_5}\,$}
\newcommand{\ud}{\mathrm{d}}
\newcommand{\ub}{\mathrm{b}}
\newcommand{\ybr}{y_\ub}
\newcommand{\Xbr}{X_\ub}
\newcommand{\kappafive}{\kappa_{_5}}

\newcommand{\tension}{\mathcal{T}}

\newcommand{\bk}{\mathbf{k}}
\newcommand{\bn}{\mathbf{n}}
\newcommand{\br}{\mathbf{r}}
\newcommand{\bp}{\mathbf{p}}
\newcommand{\bx}{\mathbf{x}}
\newcommand{\bSi}{\mathbf{\Sigma}}
\newcommand{\xx}{{\rm x}}
\newcommand{\bxp}{\mathbf{x}_{\parallel}}
\newcommand{\bpp}{\mathbf{p}_{\parallel}}
\newcommand{\pp}{p_{\perp}}
\newcommand{\microm}{\,\mathrm{\mu m}}
\newcommand{\planck}{{_\mathrm{P\negthinspace\ell}}}

\newcommand{\bea}{\begin{eqnarray}}
\newcommand{\eea}{\end{eqnarray}}
\newcommand{\be}{\begin{equation}}
\newcommand{\ee}{\end{equation}}
\newcommand{\ra}{\rightarrow}
\newcommand{\Om}{\Omega}
\newcommand{\om}{\omega}
\newcommand{\al}{\alpha}
\newcommand{\de}{\delta}

\newcommand{\ep}{\epsilon}
\newcommand{\ka}{\kappa}

\newcommand{\Si}{\Sigma}
\newcommand{\La}{\Lambda}
\newcommand{\Ga}{\Gamma}
\newcommand{\dd}{\partial}
\newcommand{\cd}{\cdot}
\newcommand{\BB}{\mathcal{B}}
\newcommand{\CC}{\mathcal{C}}
\newcommand{\EE}{\mathcal{E}}
\newcommand{\MM}{\mathcal{M}}
\newcommand{\HH}{\mathcal{H}}
\newcommand{\lap}{\Delta}

\def\Ze{\mathbb Z}      

\newcommand{\si}[1]{{\scriptscriptstyle{#1}}}

\begin{document}

\title{Tachyonic perturbations in \ads orbifolds}
\author{Cyril Cartier}
\email{cyril.cartier@physics.unige.ch}
\author{Ruth Durrer}
\email{ruth.durrer@physics.unige.ch}
\affiliation{D\'epartement de Physique Th\'eorique, Universit\'e de
Gen\`eve, 24 quai Ernest Ansermet, 1211 Gen\`eve 4, Switzerland.}

\date{\today}

\begin{abstract}
We show that scalar as well as vector and tensor metric
perturbations in the Randall-Sundrum II braneworld allow
normalizable tachyonic modes, \IE possible instabilities. These
instabilities require nonvanishing initial anisotropic stresses
on the brane. We show with a specific example that within the
Randall-Sundrum II model, even though the tachyonic modes are
excited, no instability develops.  We argue, however, that in the
cosmological context instabilities might in principle be present.
We conjecture that the tachyonic modes are due to the singularity
of the orbifold construction. We illustrate this with a simple but
explicit toy model.
\end{abstract}

\maketitle

\section{Introduction}
Already at the beginning of the last century, the idea that our
universe may have more than three spatial dimensions has been explored
by Nordstr\"om~\cite{Nordstrom:1914}, Kaluza~\cite{Kaluza:1921tu} and
Klein~\cite{Klein:1926tv}. Since superstring theory, the most
promising candidate for a theory of quantum gravity, is consistent
only in ten space-time dimensions (11 dimensions for M-theory) these
ideas have been revived in recent
years~\cite{Polchinski:1998rq,Polchinski:1998rr,Horava:1996qa}. It has
also been found that string theories naturally predict lower
dimensional ``branes" to which fermions and gauge particles are
confined, while gravitons (and the dilaton) propagate in the
bulk~\cite{Antoniadis:1990ew,Polchinski:1995mt,Lukas:1998qs}.

Recently it has been emphasized that relatively large extra-dimensions
(with typical length $L \simeq \microm$) can ``solve'' the hierarchy
problem: The effective four-di\-men\-sional Newton constant given by
$G_4 \approx G/L^n$ can become very small even if the fundamental
gravitational constant $ G \simeq m_{\planck}^{-(2+n)}$ is of the
order of the electroweak
scale~\cite{Arkani-Hamed:1998rs,Arkani-Hamed:1998nn,%
Antoniadis:1998ig,Randall:1999ee}. Here $n$ denotes the number of
extra dimensions.  It has also been shown that extra dimensions may
even be infinite if the geometry contains a so-called ``warp
factor". An especially attractive model of this type, where the bulk
is a five-dimensional anti-de Sitter (\ads) space has been developed by
Randall and Sundrum~\cite{Randall:1999vf}. This is the model which we
discuss in this work; we shall call it RSII in what follows.

The size of the extra dimensions is constrained by the requirement of
recovering usual four-dimensional Newton's law on the brane, at
least on scales tested by
experiments~\cite{Long:2003dx,Uzan:2000mz,Allen:2000ih}.

Models with finite extra dimensions always have to invoke some
nongravitational interaction in order to stabilize the graviscalar
(which is equivalent to the radion)~\cite{Durrer:2003rg}. However, in
the case of noncompact warped extra dimensions, it can happen that
this mode is not normalizable and therefore cannot be excited.  This
is precisely what happens in the RSII model.

Therefore, there is justified hope that, for suitable parameters, this
model can reproduce four-dimensional gravity without invoking
\textit{ad hoc} additional interactions. However, we show in this
paper that the gravitational sector coupled to a brane with
nonvanishing anisotropic stresses does have negative mass modes. We
argue that, on the linearized level, these instabilities are not
relevant for the Randall-Sundrum model, but they may be devastating in
the cosmological context where the brane is moving.

The tachyonic modes are absent if there are no anisotropic
stresses. Furthermore, if anisotropic stresses remain small, they
cannot develop an instability. As we shall show, this is the case
for the RSII model since there, to first order, anisotropic
stresses evolve like in Minkowski space-time and hence remain small
(if their Minkowski evolution is not already unstable). In the
cosmological context, however, this is no longer true and large
deviations from homogeneity and isotropy may in principle develop.

The outline of the paper is as follows: In the next section, the
perturbation theory on RSII is briefly introduced and the relevant
perturbation equations are given. We present the solutions to the
bulk perturbation equations and the junction conditions for
tensor, vector and scalar modes. We pay particular attention to
the tachyonic modes which are new and represent a possible
instability. In Section~\ref{Liou} we discuss the simple case of
free-streaming, relativistic particles and show that they induce
negative mass modes. We explicitly solve the equations for
the RSII background and see that no instability is induced in this
case. We then argue that, in principle, this behavior may change
in a cosmological setting.

In Sec.~\ref{RSmod}, we present a simple $3+1$ dimensional
Minkowski orbifold who's bulk modes exhibit the same instability as
the \ads orbifold. We explicitly re-construct the instability
from the retarded Green's function, showing that it is causal. In this
toy model, instabilities develop due to nonlinear couplings. A final
section is devoted to some conclusions.


\section{Perturbations of the RSII model}
\label{s:perts}
Our universe is considered to be a $3$-brane embedded in
five-dimensional anti-de Sitter space-time,
\begin{equation}
\label{eq:metric}
 ds^2
 = g_{\si{AB}} dx^{\si{A}} dx^{\si{B}}
 = \frac{L^2}{y^2} \left[-dt^2 + \delta_{ij} dx^i dx^j + dy^2\right]~.
\end{equation}
Capital Latin indices $A,B$ run from $0$ to $4$ and lower case Latin
indices $i,j$ from $1$ to $3$. Four-dimensional indices running from
$0$ to $3$ will be denoted by lower case Greek letters. Anti-de Sitter
space-time is a solution of Einstein's equations with a negative
cosmological constant $\Lambda$,
\begin{equation}
 \label{eq:einstein}
 G_{\si{AB}} + \Lambda g_{\si{AB}} = 0~.
\end{equation}
The curvature radius $L$ is given by
\begin{equation}
 \label{eq:Lambda}  L^2 = -\frac{6}{\Lambda}~.
\end{equation}

In braneworld models one often uses Gaussian normal coordinates.
For anti-de Sitter space these are given by
the transformation $L^2/y^2 = \exp{(-2 \varrho /L)}$. The metric then
takes the form
\begin{equation}
 \label{eq:adsrs}
 ds^2
 = g_{\si{AB}} dx^{\si{A}} dx^{\si{B}}
 = e^{-2 \rho /L} \left(-dt^2 + \delta_{ij} dx^i
   dx^j \right) + d\rho^2~.
\end{equation}

We now introduce one single brane at $y=\ybr=L$ (or equivalently $\rho
=0$) and replace the ``left hand side'', $0<y<L$, of \ads by a second
copy of the ``right hand side''. We use the superscripts ``${\si >}$''
and ``${\si <}$'' for the bulk sides with $y > \ybr$ and $y < \ybr$,
respectively. In terms of the coordinate $y$, the value of $y$
decreases continuously from $\infty$ to $L$ and then jumps to $-L$
over the brane whereafter it continues to decrease. At the brane
position, $\ybr^{\si >}=L, ~ \ybr^{\si <} = -L$, the metric function
$(L/y)^2$ has a kink. The advantage of the coordinate $\rho$
introduced in Eq.~\eqref{eq:adsrs} is that the variable $\rho$ does
not jump, but the metric function in the presence of a brane becomes
$\exp(-2|\rho|/L)$.

The Einstein equations at the brane position are singular; they
contain a Dirac-delta function. To avoid this, one can integrate them
over the brane which leads to the so-called junction
conditions~\cite{Lanczos:1924,Sen:1924,Darmois:1927,Israel:1966} at
the brane position. These read~\cite{Misner:1970aa}
\begin{equation}\label{eq:jump}
 K_{\mu \nu}^{\si >} - K_{\mu \nu}^{\si <}
 = \kappafive\left( S_{\mu \nu} - \frac{1}{3} S q_{\mu \nu} \right)
 \equiv \kappafive \widehat{S}_{\mu \nu}~,
\end{equation}
where $S_{\mu \nu}$ is the energy-momentum tensor on the brane with
trace $S$, and
\begin{equation}
 \kappafive \equiv 6\pi^2 G_5=\frac{1}{M_5^{3}}~.
\end{equation}
$M_5$ and $G_5$ are the five-dimensional (fundamental) reduced
Planck mass and Newton constant, respectively. $K_{\mu\nu}$ is the
extrinsic curvature of the brane and $q_{\mu\nu}$ is the induced
metric on the brane. Equation~\eqref{eq:jump} is usually referred
to as the second junction condition. The first junction condition
simply states that the induced metric, the first fundamental form,
\begin{equation} \label{eq:firstfund}
q_{\mu\nu} = e^{\si{A}}_\mu e^{\si{B}}_\nu g_{\si{AB}}~,
\end{equation}
be continuous across the brane. Here the vectors $e_\nu$ are
tangent to the brane. In more detail, if we parametrize the brane
by coordinates $\left(z^\mu\right)$ and its position in the bulk
is given by functions $\Xbr^{\si{A}}(z^\mu)$, the vectors $e_\nu$
are defined by
\begin{equation}
 e_\mu^{\si{A}} = \dd_\mu \Xbr^{\si{A}}(z) ~.
\end{equation}
Denoting the brane normal by $n$, we have $g_{\si{AB}}e^{\si{A}}_\mu
n^{\si{B}}~=~0$. The extrinsic curvature can be expressed purely in
terms of the internal brane
coordinates~\cite{Deruelle:2000yj,Mukohyama:2001yp},
$K~=~K_{\mu\nu}dz^\nu dz^\mu$, with
\begin{equation}
\label{eq:extusefull}
K_{\mu \nu} = -\frac{1}{2} \left[g_{\si{AB}}\left(e^{\si{A}}_\mu
\partial_\nu n^{\si{B}} + e^{\si{A}}_\nu  \partial_\mu
n^{\si{B}}\right) + e^{\si{A}}_\mu e^{\si{B}}_\nu n^{\si{C}}
g_{\si{AB},\si{C}} \right]~.
\end{equation}

In the case we are interested in, the background space-time
consists of two copies of the part of \ads with $|y| \geqslant
|\ybr|=L$. We actually let the coordinate $y$ jump from $y=L$ to
$y=-L$ across the brane. Since the metric is symmetric in $y$, the
first junction condition is trivially fulfilled. The second
fundamental form is proportional to the induced metric,
$K_{\mu\nu} = \pm L^{-1}q_{\mu\nu}$, hence the energy-momentum
tensor on the brane is a pure brane tension $\tension$,
$S_{\mu\nu} = -\tension q_{\mu\nu}$. With
\begin{equation}
 K_{\mu\nu}^{\si >} - K_{\mu\nu}^{\si <}
 = \left[K_{\mu\nu}\right]
 = 2K_{\mu\nu}^{\si >} ~,
\end{equation}
the second junction condition becomes
\begin{align}
 &\left.[K_{00}]\right|_{\ybr}
    = -\frac{2}{L}
    = -\frac{1}{3}\kappafive \tension~,\label{eq:2ndback} \\
 & \left.[K_{ii}]\right|_{\ybr}
    = \frac{2}{L}
    = \frac{1}{3}\kappafive \tension~.
\end{align}
This leads to the well-known RS-fine-tuning condition,
\begin{equation}
 \label{eq:fine}
 -\La = \frac{6}{L^2} = \frac{1}{6}\kappafive^2 \tension^2~.
\end{equation}

The most general perturbation of the \ads metric \eqref{eq:metric} is
of the form
\begin{align}
 \label{eq:pertgen}
 ds^2
 &= g_{\si{AB}} dx^{\si{A}} dx^{\si{B}} \nonumber \\
 &= \frac{L^2}{y^2} \left[-(1+2\Psi)dt^2
     -4\Si_i dt dx^i-4\BB dt dy  \nonumber \right. \\
 &\quad  \left. +\left((1-2\Phi)\delta_{ij} +2H_{ij}\right)\ud
     x^i dx^j +4\Xi_idx^idy \nonumber \right. \\
 &\quad \left. +(1+2\CC) dy^2 \right]~.
\end{align}
Here $\Si_i$ and $\Xi_i$ are divergenceless vectors and $H_{ij}$
is a divergenceless, traceless tensor. It is easy to show that
there exists one fully specified gauge in which the perturbation
variables take this form; vectors have no ``scalar component'' and
tensors have neither a vector nor a scalar component. We call this
the generalized longitudinal gauge (see
also~\cite{vandeBruck:2000ju,Riazuelo:2002mi}). We shall use it in
the following. Within linear perturbation theory, the variables
with different spin, the tensor $H_{ij}$, the vectors $\Si_i$ and
$\Xi_i$, and the scalars $\Psi,\Phi, \BB, \CC$ do not couple. We
can therefore study the perturbations of each type separately. We
shall do so in the next subsections. There we write down the
perturbed Einstein equations for a fixed Fourier-mode $\bk$ for
which we have $\bk\cdot\mathbf{\Si} = \bk\cdot \mathbf{\Xi} =
k^iH_{ij} = 0$. We do not perform a Fourier decomposition in time.

We want to study the perturbations in an empty bulk with possible
perturbations on the brane. The five dimensional Einstein equation
implies the perturbation equations in the bulk,
\begin{equation}
 \de G_{\si{AB}}
 = -\La\de g_{\si{AB}}~,
\label{eq:pertEi}
\end{equation}
and the junction conditions at the brane,
\begin{equation}\label{eq:junc}
 2\de K_{\mu\nu} =\ka_5\de\widehat{S}_{\mu\nu}~.
\end{equation}
We first discuss tensor perturbations. As we shall see, the
homogeneous equations reduce to the same Bessel equations for all
three types of perturbations (see also~\cite{Deruelle:2000yj}).

\subsection{Tensor perturbations}
In this paragraph we first discuss the simplest case, the tensor
perturbation equations. We write them down for a fixed Fourier-mode
$\bk$ and determine their solutions. We consider only $H_{ij}\neq
0$. For this case, Eq.~\eqref{eq:pertEi} reduces to
\begin{equation}
\label{eq:tenEi}
\left(\dd_t^2 +k^2 -\dd_y^2 + \frac{3}{y}\dd_y \right)H_{ij} = 0~.
\end{equation}
For a given polarization, $H_{\si\bullet} = H_+$ or $H_{\si\bullet} =
H_{\times}$, we make the ansatz $H_{\si\bullet} =f(t)g(y)$ leading to
\begin{equation}
\label{eq:tent}
 \frac{\dd_t^2f}{f} +k^2
 = \frac{(\dd_y^2 - \frac{3}{y}\dd_y)g}{g}
 =Z~,
\end{equation}
where $Z$ is an arbitrary separation constant. The behavior of the
solutions to these equations depends strongly on the sign of $Z$. If
$Z=-m^2$ is negative, we obtain
\begin{align}
\label{eq:ten+}
f &= \exp(\pm i t\sqrt{m^2+k^2}) \equiv  \exp(\pm i \om t)~,\\
g &= N(my)^2\times \left\{
     \begin{array}{l}
       J_2(my)~, \\[1mm]
       Y_2(my)~.
     \end{array} \right.
\end{align}
Here $J_\nu$ and $Y_\nu$ denote the Bessel functions of order
$\nu$. They are oscillating and decaying. They are ``$\de$-function
normalizable'' perturbations like harmonic waves in flat space, in the
sense that $H_m=fg$ satisfies~\cite{Csaki:2000fc,Bozza:2001xt}
\begin{equation}
 \int_0^\infty H_mH_{m'} \frac{dy}{m^2y^3} \propto m\de(m-m') ~.
 \label{e:normalisation}
\end{equation}
These are just the ordinary gravity modes of four-dimensional mass $m$
without a mass gap which are discussed in the original RS
paper~\cite{Randall:1999vf}. However, if $Z=-m^2$ is positive, the
solutions take the form
\begin{align}
\label{eq:ten-}
 f &= \exp(\pm  t\sqrt{Z-k^2})  = \exp(\pm  t\om)~,\\
 g &= N(|m|y)^2\times  \left\{ \begin{array}{l}
   K_2(|m|y)~, \\[1mm]
   I_2(|m|y)~. \end{array} \right.
\end{align}
Here $K_\nu$ and $I_\nu$ are the modified Bessel functions of order
$\nu$. The second case $g \propto I_2$ grows exponentially in
$y$. This is not normalizable and therefore cannot represent a
physical, small perturbation. However, the mode $K_2$ decays
exponentially and is normalizable and small for sufficiently small
initial amplitudes. However, even with arbitrary small initial data
this mode grows exponentially in time for sufficiently small wave
numbers, $k^2<-m^2$~; it is a tachyonic instability.

To have a complete solution to the perturbation equations we need to
discuss the boundary conditions at the brane, \IE the junction
conditions.

A short computation shows that the nonvanishing components of the
extrinsic curvature tensor perturbations are in our case
\begin{align}
  \left. \de K_{ij} \right|_{\ybr}
   & = \left.\left(\frac{2}{L}H_{ij}-\dd_y H_{ij}\right)\right|_{\ybr}~,
    \quad\mbox{hence}  \nonumber \\  \label{eq:2ndten}
 - \left. 2(\dd_y H_{ij}) \right|_{\ybr}
   & = \kappafive \Pi^{(T)}_{ij}~,
\end{align}
where $\Pi^{(T)}$ are tensor-type anisotropic stresses on the brane.

Let us first consider the homogeneous case $\Pi^{(T)}\equiv0$.  For
$m^2>0$, the solutions are of the form
\begin{equation}
 H = \exp(\pm i\om t)(my)^2\left[ AJ_2(my) +BY_2(my)\right]  ~.
\end{equation}
The junction condition~\eqref{eq:2ndten} then requires
\begin{equation}
 B = -A\frac{J_1(mL)}{Y_1(mL)} \simeq \frac{\pi}{4}(mL)^2A ~,
\end{equation}
where the last expression is a good approximation for $mL\ll 1$. This
is precisely the result of Randall and
Sundrum~\cite{Randall:1999vf}. This is not modified even if we allow
for the negative mass modes, $Z=-m^2 >0$, because a physical solution
has to be of the form
\begin{equation}\label{eq:tachten}
  H = C\exp(\pm t\sqrt{Z-k^2})(|m|y)^2K_2(|m|y)~,
\end{equation}
and since $K_1$ has no zero, the junction condition \eqref{eq:2ndten}
requires $C=0$.

But in a realistic brane universe, $\Pi^{(T)}$ is not exactly zero. In
cosmology, it is just typically a factor of 10 smaller than other
perturbations of the energy-momentum tensor on the brane. We therefore
can not set $C\equiv 0$. However, as long as $\Pi^{(T)}$ remains
small, we do not expect the unstable modes to be present; hence we
expect $C(k,m)=0$ for $k^2<-m^2$. In the next section, we shall show
in a specific example that this is indeed the case in RSII, where the
brane is Minkowski space and {\em to linear order} the anisotropic
stress follows its background equation of motion. In the cosmological
context, however, the evolution of $\Pi^{(T)}$ contains $H$ which then
can in principle feed the instability.

One may ask, whether the unstable $K$ mode is a consequence of the
thin brane limit. However, it is clear that in a thick
brane, the kink of the $K$ mode at the brane position will simply be
replaced by a rapid but gradual transition. The important point
is that the presence of the brane cuts off the inner part $0<y<L$ of
AdS$_5$ on which the $K$ mode would not be normalizable, and replaced
it by a second copy of $L<y<\infty$, rendering the $K$ mode
normalizable in the bulk. It is this orbifold contruction and not the
limit of an infinitely thin brane which leads to the instability.

It is interesting to note that the $K$ mode decays exponentially in
$y$ direction; it does not propagate into the bulk, but moves along
the brane. However, due to its negative mass square, $m^2<0$, it is a
tachyon and therefore provokes an instability, and its amplitude grows
exponentially in time for sufficiently small wave numbers, $k^2<-m^2$. 

In Sec.~\ref{RSmod} we discuss a toy model, where we see
explicitly how the exponential growth of $H$ can trigger an
exponential growth of quantity on  the brane it couples to, when
nonlinear effects are taken into account. In that sense it is
not clear that the RSII model is safe from this instability
once non--linear effects are taken into account.

The discovery of these tachyonic ``modes''\footnote{Since the new
``K modes'' are not free, they have to vanish if anisotropic stresses
are absent, they are not modes in the usual sense of the term:
free solutions to some linear hyperbolic equation. We shall
nevertheless use this term here, committing a slight abuse of
language.} is already the main result of this work. In the next
subsections we simply repeat our analysis for vector and scalar
perturbations and show that exactly the same behavior is found there.

\subsection{Vector perturbation}

In this section we follow closely Ref.~\cite{Ringeval:2003na} where
the problem of vector perturbations has been considered in the
cosmological context.

The bulk Einstein equations for a mode $\bk$ of the vector
perturbations $\Si$ and $\Xi$ are
\begin{align}
 \left(\dd_y^2-\frac{3}{y}\dd_y\right)\Si
 &= \left(\dd_t^2+k^2\right)\Si~, \label{eq:bulkvecSi}\\
 \left(\dd_y^2-\frac{3}{y}\dd_y+\frac{3}{y^2}\right)\Xi
 &=\left(\dd_t^2+k^2\right)\Xi~,\label{eq:bulkvecXi}\\
 \left(\dd_y -\frac{3}{y}\right)\Xi
 &= - \dd_t\Si~,\label{eq:bulkvecconst}
\end{align}
where the spatial index on $\Si$ and $\Xi$ has been omitted.  The
constraint equation Eq.~\eqref{eq:bulkvecconst} fixes the relative
amplitudes of $\Si$ and $\Xi$, showing that there is only one
independent vector perturbation in the bulk (the ``graviphoton'').
One can check that these equations are consistent, \EG with the master
function approach of Ref.~\cite{Mukohyama:2000ui}.

As in the tensor case, one can solve the equations with a separation
ansatz. For a negative separation constant, $Z=-m^2<0$, one obtains
the expected oscillatory modes,
\begin{align}
 \Si &= \exp(\pm i\om t)(my)^2\left[A J_2(my) + BY_2(my) \right]~, \\
 \Xi &= \frac{\pm i\om}{m}  \exp(\pm i\om t)(my)^2\left[A
 J_1(my) +  BY_1(my) \right]~,
\end{align}
where $\om =\sqrt{m^2+k^2}$. These solutions have been found in
Ref.~\cite{Bridgman:2000ih}. For a positive separation constant
$Z=-m^2> 0$, we obtain again tachyonic solutions. Like in the tensor
case, the solution containing the modified Bessel function $I_\nu$
cannot be accepted as it is exponentially growing and thus represents
a non-normalizable perturbation. However, the $K_\nu$-solution is
exponentially decaying and therefore perfectly acceptable. For
tachyonic vector perturbations with $\om = \sqrt{Z-k^2}$ we have
\begin{align}
 \Si &= C\exp(\pm \om t)(|m|y)^2K_2(|m|y)~,  \\
 \Xi &= \frac{\pm \om}{|m|} C\exp(\pm \om t)(|m|y)^2K_1(|m|y)~.
\end{align}
For large enough scales, $-m^2>k^2$, these solutions can grow
exponentially.

Again, the boundary conditions at the brane relate the
perturbations to the brane energy-momentum tensor. For the
energy-momentum tensor on the brane, the vector degrees of freedom
are defined according to
\begin{equation}
 S_{\mu\nu}
 = \begin{pmatrix}0 & 2V_j \cr 2V_i & 4\Pi^{(V)}_{ij}\end{pmatrix}
   -\tension q_{\mu\nu}~,
 \label{e:T_vector}
\end{equation}
where $V_i$ and $\Pi^{(V)}_i$ are divergence-free vectors fields
and $\Pi^{(V)}_{ij}\equiv\frac{1}{2}[\dd_j \Pi^{(V)}_i+\dd_i
\Pi^{(V)}_j]$. The first junction condition simply requires that
$\Si$ be continuous at the brane, which it is since the (modified)
Bessel functions of even index are even functions. The second
junction condition results in (for a detailed derivation,
see~\cite{Ringeval:2003na})
\begin{align}
 \left. (\dd_t\Xi+\dd_y \Sigma)\right|_{\ybr} &= 
\kappafive V~, \label{eq:jv1}\\
\left. \Xi\right|_{\ybr} &= \kappafive \Pi^{(V)}~,\label{eq:jv2} \\
 \dd_t V &=-k^2 \Pi^{(V)}~.
\end{align}
The last equation follows from \eqref{eq:jv1} and \eqref{eq:jv2} and
the bulk equations of motion
\eqref{eq:bulkvecSi}--\eqref{eq:bulkvecconst}. It represents momentum
conservation on the brane, which is guaranteed as long we have
vanishing energy flux off the brane and $\Ze_2$--symmetry.

Like for tensor perturbations, we first consider homogeneous
solutions, setting $\Pi^{(V)}\equiv V \equiv 0$. This requires
$\Xi(|m|L)=0$, hence
\begin{align}
 B &= -A\frac{J_1(mL)}{Y_1(mL)} \quad\mbox{ for } m^2>0~,
 \label{e:B-vector-onbrane}\\
 C  &\equiv 0  \quad \mbox{ for } m^2<0~.
 \label{e:C-vector-onbrane}
\end{align}
Equation~\eqref{eq:jv1} is then identically satisfied.

However, it seems more realistic to allow for a small but
nonvanishing anisotropic stress contribution $\Pi^{(V)}$ and
corresponding vorticity $V$. In this case, again, solutions with
$C\neq 0$ may exist, and small initial data could lead to 
exponential growth like for tensor perturbations.

Using the normalization condition \eqref{e:normalisation} for the
$m=0$ mode of the variable $\Xi\propto y$ (this is the one which
enters as dynamical variable in the perturbed action;
see~\cite{Csaki:1999jh}), one finds that contrary to the tensor case,
the vector zero-mode is not normalizable. Therefore, on the brane
there is only the ordinary massless spin-$2$ graviton, but there are an
infinity of massive spin-$2$ and spin-$1$ particles (the modes discussed
here, with $m\neq 0$).

\subsection{Scalar perturbation}
We now discuss the most cumbersome, the scalar sector. Scalar-type
metric perturbations in the bulk are of the form
\begin{align}\label{eq:scal}
 ds^2
 &= \frac{L^2}{y^2}
     \left[-(1+2\Psi)dt^2  -4\BB dt dy \right.  \nonumber  \\
 &\quad \left.+(1-2\Phi)\delta_{ij}dx^i dx^j
    +(1+2\CC) dy^2 \right]~.
\end{align}

The bulk Einstein perturbation equations for the mode $\bk$ become,
after some manipulations and introducing the combination
$\Ga\equiv\Phi +\Psi $,
\begin{align}
 \Phi - \Psi &=\CC~,   \label{eq:psi}\\
 \left(\dd_y^2 -\frac{3}{y}\dd_y\right)\Ga
 &= \left(\dd_t^2 +k^2\right)\Ga ~, \label{eq:ga}\\
 \left(\dd_y^2 -\frac{3}{y}\dd_y + \frac{4}{y^2}\right)\CC
 &= \left(\dd_t^2 +k^2\right)\CC  ~, \label{eq:c}\\
 \dd_y\Phi+ \left(\dd_y -\frac{3}{y}\right)\CC
 &= -\dd_t\BB ~,\label{eq:b}\\
 \frac{3}{y}\left(\dd_y-\frac{2}{y}\right)\CC
 &= 3\dd^2_t\Phi+k^2(\Phi +\CC)~,\label{eq:10}\\
 3\dd_t\left(\dd_y\Phi - \frac{1}{y}\CC\right)
 &= k^2\BB~, \label{eq:blast}\\
 \dd_t\left(2\Phi-\CC\right)
 &= \left(\dd_y -\frac{3}{y}\right)\BB~. \label{eq:last}
\end{align}

Clearly these equations are not all independent;
Eqs.~\eqref{eq:blast} and \eqref{eq:last} are identically satisfied
if Eqs.~\eqref{eq:psi}--\eqref{eq:10} are. The solutions are
obtained as for tensor and vector perturbations. For a negative
separation constant $Z = -m^2<0$, we obtain ($\om
=\sqrt{m^2+k^2}$)
\begin{align}
 \Ga &= \exp(\pm i\om t)(my)^2\left[A'J_2(my) + B'Y_2(my) \right]~, \\
 \CC &= \exp(\pm i\om t)(my)^2\left[AJ_0(my) + BY_0(my) \right] ~,\\
 \Phi &= \frac{1}{2}\exp(\pm i\om t)(my)^2
          \left[A'J_2(my)+ B'Y_2(my) \right.  \nonumber \\
      & \qquad \left. +AJ_0(my) +  BY_0(my) \right]~, \\
 \Psi &= \frac{1}{2}\exp(\pm i\om t)(my)^2
          \left[A'J_2(my)+ B'Y_2(my)\right. \nonumber \\
      & \qquad\left. - AJ_0(my) -  BY_0(my) \right]~, \\
 \BB &= \frac{\pm im^3y^2}{2\om} \exp(\pm i\om t)
        \left[(A'-3A)J_1(my)\right. \nonumber \\
     & \qquad \left.+(B'-3B)Y_1(my) \right]~, \\
 A' &= 3A\frac{m^2}{m^2+2\om^2}~, \quad
 B' = 3B\frac{m^2}{m^2+2\om^2}~.
\end{align}

For a positive separation constant, $Z=-m^2 > 0$, we find
($\om=\sqrt{Z-k^2}$)
\begin{align}
 \Ga  &= \exp(\pm \om t)(|m|y)^2C'K_2(|m|y)~, \\
 \CC  &= \exp(\pm \om t)(|m|y)^2CK_0(|m|y)~,\\
 \Phi &= \frac{1}{2}\exp(\pm\om t)(|m|y)^2  
         \left[C'K_2(|m|y)+ CK_0(|m|y) \right] ~,\\
 \Psi &= \frac{1}{2}\exp(\pm \om t)(|m|y)^2 
         \left[C'K_2(|m|y) - CK_0(|m|y) \right]~, \\
 \BB  &= \frac{\pm |m|^3y^2}{2\om} \exp(\pm \om t)[C'+3C]K_1(|m|y)~, \\
  C'  &= -3C\frac{|m|^2}{|m|^2+2\om^2}~,
\end{align}
where we have already used that the $I$ mode is not normalizable and
therefore cannot contribute. Like for vector and tensor perturbations,
we find again  tachyonic solutions with $m^2<0$ which represent an
exponential instability for sufficiently small wave number $k$ (large
scales).

Determining the boundary conditions via the first and second
junction conditions now requires a bit more care. Since we have
already fully specified our coordinate system by the adopted
choice of perturbation variables, we must allow for brane bending.
We cannot fix the brane at $\ybr=L$, but we must allow for
$\ybr^{\si >}=L+\EE$ and $\ybr^{\si <}=-L-\EE$, respectively. The
antisymmetry $\ybr^{\si <}=-\ybr^{\si >}$ is an expression of
$\Ze_2$ symmetry. The introduction of the new perturbation
variable $\EE(z^\mu)$ describing brane bending affects the
first and second fundamental forms. From Eq.~\eqref{eq:firstfund},
we obtain $q_{\mu\nu}=g_{\mu\nu}$ to first order, which implies
that $\Phi$ and $\Psi$, hence $\CC$, have to be continuous. At the
brane position, the perturbed components of the extrinsic
curvature~\eqref{eq:extusefull} are
\begin{align}
 \de K_{00}
   &= \frac{1}{L}\left[\Phi-3\Psi+ 2\frac{\EE}{L}\right]
      +\dd_y \Psi-2\dd_t \BB+\dd_t^2 \EE~,
      \label{e:K00-scalar} \\
 \de K_{0j}
   &= \dd_j\left(\dd_t \EE- \BB \right)~,
      \label{e:K0j-scalar} \\
 \de K_{ij}
   &= \left[\frac{1}{L}\left(\Psi -3\Phi -2\frac{\EE}{L}\right)
      +\dd_y \Phi\right]\delta_{ij}+\dd_i \dd_j \EE~.
      \label{e:Kij-scalar}
\end{align}
For the energy-momentum tensor on the brane, we parametrize the
$4$ degrees of freedom according to
\begin{equation}
 S_{\mu\nu}
 =  \begin{pmatrix}\rho & v_i \cr v_j & P\delta_{ij}
                   +\Pi^{(S)}_{ij} \end{pmatrix}
    -\tension q_{\mu\nu}~,
 \label{e:T_scalar}
\end{equation}
where $v_i \equiv \dd_i v$ and $\Pi^{(S)}_{ij} \equiv \left(\dd_i
\dd_j - \frac{1}{3}\Delta \delta_{ij}\right)\Pi^{(S)}$. With
Eqs.~\eqref{e:K00-scalar}--\eqref{e:Kij-scalar}, the second
junction condition reads
\begin{align}
 \frac{1}{\tension}\left(2\rho + 3P\right)
 &= \left. [\Phi\! -\! \Psi\! +\! L\dd_t\left(\dd_t\EE-2\BB\right)\!
     + \! L\dd_y \Psi] \right|_{\ybr} ~, \label{e:dK00} \\
 \frac{3}{\tension L}v
 &=\left. [ \dd_t\EE-\BB] \right|_{\ybr}~, \label{e:dK0j} \\
 \frac{3}{\tension L}\Pi^{(S)}
 &= \EE~, \label{e:dKij}\\
 \frac{1}{\tension}\left[\rho - \lap\Pi^{(S)}\right]
 &= \left. [\Psi-\Phi +L\dd_y\Phi] \right|_{\ybr} ~. \label{e:dKii}
\end{align}
Combining the time derivative of Eq.~\eqref{e:dK0j} with
Eqs.~\eqref{eq:b}, \eqref{e:dK00} and \eqref{e:dKii}, we obtain
momentum conservation on the brane,
\begin{equation}
 \dd_t v = \frac{2}{3}\Delta \Pi^{(S)}+P~.
 \label{e-scalar-const-dv}
\end{equation}
Similar manipulations imply energy conservation on the brane,
\begin{equation}
 \dd_t\rho = \lap v~.
\end{equation}

Like for tensor and vector perturbations, we first look for
solutions with vanishing brane matter. Setting $\Pi^{(S)}\equiv
\rho\equiv P\equiv v \equiv 0$ forbids brane bending, $\EE=0$.
Then Eq.~\eqref{e:dK0j} implies $\BB(mL)=0$, thus
\begin{align} \label{e:BB'-scalar-onbrane}
 B'-3B &= -(A'-3A)\frac{J_1(mL)}{Y_1(mL)}
 \quad\mbox{ for } m^2>0~, \\
 C'+3C &= 0  \quad \mbox{ for } m^2<0~.
\end{align}
The other equations are satisfied if we require separately
\begin{align}
 \frac{B}{A} &= \frac{B'}{A'}= -\frac{J_1(mL)}{Y_1(mL)}
 \quad\mbox{ for } m^2>0~,
 \label{e:B-scalar-onbrane}\\
 C  &\equiv 0  \quad \mbox{ for } m^2<0~.
 \label{e:C-scalar-onbrane}
\end{align}
Equations~\eqref{e:BB'-scalar-onbrane} and \eqref{e:B-scalar-onbrane}
are of course equivalent.

As for vector perturbations, the $m=0$ scalar mode is not
normalizable. Like for tensor and vector perturbations, we have found
``scalar gravitons'' which appear on the brane as massive
particles. If the brane matter is unperturbed, only oscillating
$m^2>0$ solutions are possible. However, if we allow for nonvanishing
matter perturbations on the brane, we can have $C\neq 0$ and the
tachyonic modes $m^2<0$ can appear exactly like in the tensor and
vector sectors.

It is not surprising that the same instability appears in the scalar,
vector and tensor sectors, because all modes describe the same bulk
particle, the five-dimensional graviton.

\section{Tachyonic modes from freely propagating relativistic particles}
\label{Liou}

\subsection{The Liouville equation in 4 dimensions}
As long as $\Ze_2$ symmetry is satisfied, we expect free
(collisionless) particles on the brane to move along
brane geodesics. Their one-particle distribution function therefore
obeys the Liouville equation.  Here we sketch a derivation of
perturbation to the $4$D Liouville equation. Many more details can be
found, \EG in~\cite{Durrer:1993db} or~\cite{Durrer:1998rw}.

An ensemble of freely propagating particles on the brane is described
by the Liouville equation,
\begin{equation} \label{eq:liou1}
\left[ p^\mu\dd_\mu  - \Ga^i_{\al\beta}p^\al p^\beta
  \frac{\dd}{\dd p^i} \right]f =L_{X_g}f= 0~.
\end{equation}
Here $f$ is the one-particle distribution function defined on the mass
shell,
\begin{equation}
 P_M \equiv
   \left\{ (x^\mu,p^\nu) \mid g_{\mu\nu}(x)p^\mu p^\nu = -M^2 \right\}~,
\end{equation}
which we parametrize by the 7 coordinates $(x^\mu, p^i)$. The
energy $p^0$ is then determined via the mass-shell condition. The
energy-momentum tensor of the particles is given by
\begin{equation} \label{eq:em}
 T^{\mu\nu} = \int \frac{d^3p \sqrt{|g|}}{ p^0} p^\mu p^\nu f~.
\end{equation}
To maximize anisotropic stresses, we consider relativistic particles,
$M^2=0$. In this case we obtain the simple relation
\begin{equation}
 \rho = - T^0_0 = 3P = T^i_i ~,
\end{equation}
where $\rho$ is the energy density and $P$ the ``pressure'' of the
collisionless particles. With respect to an orthonormal frame we
parametrize the particle momentum $p^i = pn^i$ and define the
``brightness''
\begin{equation}\label{eq:MM}
\MM = \frac{4\pi a^4}{3} \int dp\,p^3 f ~.
\end{equation}
Here $a$ is the cosmic scale factor which we can simply set to $a=1$
in the case of a nonexpanding background. The components of the
energy-momentum tensor are then given by integrals over the
momentum directions $\bn$. The anisotropic stress is
\begin{equation}\label{eq:PiMM}
 \Pi_{ij} = \frac{1}{4\pi a^4} \int d\Om \left( n_in_j-\frac{1}{3}
 \de_{ij} \right)\MM ~.
\end{equation}

Let us first consider tensor perturbations of the metric. Then, the
Liouville equation~\eqref{eq:liou1} for the spatial Fourier transform
of $\MM$ becomes, to first order in the metric
perturbations,~\cite{Durrer:1993db}
\begin{equation} \label{eq:liouT}
 \dot{\MM}^{(T)} +ik\mu\MM^{(T)}
 = -4a^4Pn^in^j\dot H_{ij}~,
\end{equation}
where $\mu=\widehat\bk\cdot\bn$ is the direction cosine between
$\bk$ and $\bn$, $\widehat\bk = \bk/k$. The overdot denotes the
derivative with respect to conformal time on the brane, $\eta$. We now choose
the coordinate system so that $\bk$ is in $3$-direction and
$(\theta,\varphi)$ denote the usual polar angles with
$\mu=\cos(\theta)$. In order to represent a pure tensor
perturbation, $\MM^{(T)}$ must be of the form
\begin{align}
\MM^{(T)}(\bk,\bn,\eta) &=
(1-\mu^2)\left[\MM_+(k,\mu,\eta)\cos(2\varphi) \right. \nonumber \\
\label{eq:MMT}
 & \left. \quad + \MM_{\times}(k,\mu,\eta)\sin(2\varphi)\right] ~.
\end{align}
Using this ansatz, the two modes of $H_{ij}$ decouple and
Eq.~\eqref{eq:liouT} reduces to
\begin{equation} \label{eq:liouT2}
 \dot{\MM}_{\si\bullet} +ik\mu\MM_{\si\bullet}
 = -4a^4P\dot{H}_{\si\bullet}~,
\end{equation}
where ``$\bullet$'' stands for ``$\times$'' or ``$+$''.
Decomposing the tensor-type anisotropic stress into the two
standard helicities $+$ and $\times$, we obtain
\begin{equation}\label{eq:PiT}
\Pi_{\si\bullet}^{(T)} = \frac{3}{8a^4} \int_{-1}^1
d\mu\,(1-\mu^2)^2 \MM_{\si\bullet}(k,\mu,\eta) ~.
\end{equation}

For vector perturbations one finds, correspondingly
\begin{equation} \label{eq:liouV}
\dot\MM^{(V)} +ik\mu\MM^{(V)}
 = -4a^4Pik\mu\left(\bn\cdot\bSi\right)~,
\end{equation}
where $\Si_i$ is the perturbation of $g_{0i}$ [corresponding to our
vector perturbation in Eq.~\eqref{eq:pertgen} at fixed $y$].  Using
the coordinate system where $\bk$ points in the third axis,
$\MM^{(V)}$ must be of the following form to represent a pure vector
perturbation,
\begin{align}
 \MM^{(V)}(\bk,\bn,\eta)
 &= \sqrt{1-\mu^2}\left[\MM_1(k,\mu,\eta)\cos(\varphi)\right.
    \nonumber \\
 & \left. \quad+\MM_2(k,\mu,\eta)\sin(\varphi)\right] ~.
 \label{eq:MMV}
\end{align}
Again, with this ansatz the equations for the two helicities decouple
into
\begin{equation} \label{eq:liouV2}
\dot{\MM}_{\si\bullet} +ik\mu\MM_{\si\bullet} = -4a^4Pik\mu\Si_{\si\bullet}~,
\end{equation}
and the components of the anisotropic stress potential $\Pi^{(V)}_j$ are
given by
\begin{equation}\label{eq:PiV}
\Pi_{\si\bullet}^{(V)} = \frac{-i}{2a^4} \int_{-1}^1
d\mu\,(1-\mu^2)\mu \MM_{\si\bullet}(k,\mu,\eta)~.
\end{equation}

Finally, for scalar perturbations of the metric, which are given by
the Bardeen potentials, $\Psi_\ub$ and $\Phi_\ub$ which are the
longitudinal perturbations of the induced metric $q_{\mu\nu}$ on the
brane, one finds
\begin{equation} \label{eq:liouS}
\dot\MM^{(S)} +ik\mu\MM^{(S)} = -4a^4Pik\mu\left(\Psi_\ub+\Phi_\ub\right) ~.
\end{equation}
To represent a pure scalar mode, $\MM^{(S)}$ must be independent of the
polar angle $\varphi$.  The anisotropic stress potential $\Pi^{(S)}$
is thus given by
\begin{equation}
\Pi^{(S)} =\frac{1}{4a^4} \int_{-1}^1d\mu
\left(\frac{1}{3}-\mu^2\right) \MM^{(S)}(k,\mu,\eta)~.
\end{equation}

\subsection{Solution in the RSII model}
In the RSII model, the brane is Minkowski space-time. We therefore can
simply fix $a=1$ and conformal time is identical to physical
time, $\eta=t$. Furthermore, there is no matter in the background so that
the functions $\MM$ and the pressure $P$ are both of first
order. Therefore, the right-hand sides of the perturbation
Eqs.~\eqref{eq:liouT}, \eqref{eq:liouV} and \eqref{eq:liouS} are of second
order and have to be dropped in a consistent first order treatment.
The solution is therefore of the same form in all cases,
\begin{equation}\label{eq:MMsol}
\MM = F(k,\mu,\varphi)\exp(-ik\mu t)~.
\end{equation}
Before we determine the prefactor $F$, and the resulting anisotropic
stress $\Pi_{ij}$, we Fourier transform $\MM$ with respect to
time. This yields
\begin{equation}
 \MM(k,\mu,\varphi,\om)
 = F(k,\mu,\varphi)\de(\om-k\mu)~.
\end{equation}

For \textit{tensor} perturbations,
\begin{align}
 F(k,\mu,\varphi)
 &=(1-\mu^2)\left[ \MM_+(\mu,k)\cos(2\varphi)\right. \nonumber \\
 &\quad\left.+\MM_\times(\mu,k)\sin(2\varphi)\right]~,
\end{align}
we find
\begin{equation}
 \Pi_{\si\bullet}^{(T)} =
 \left\{\begin{array}{ll}
    \frac{3}{8}\MM_{\si\bullet}(\om/k,k)
    \left(1-\frac{\om^2}{k^2}\right)^2
   & \mbox{ for }~\om^2<k^2~, \\
 0 & \mbox{ else}~.  \end{array} \right.
\end{equation}
Comparing this with Eq.~\eqref{eq:2ndten} and using $\om^2=k^2+m^2$,
we find that only modes with $-k^2\leqslant m^2<0$ are excited and
\begin{equation}
\begin{aligned}
 C_{\si\bullet}(m^2,k)
 &= \frac{-3\kappa_5}{16} F_{\si\bullet}
    \left(\tfrac{\sqrt{k^2+m^2}}{k},k\right)
    \frac{|m|^2}{L^2k^4K_1(|m|L)}~\\
 & \qquad\mbox{if} -k^2\leqslant m^2<0 ~,\\
 C_{\si\bullet}(m^2,k) & = 0 \quad\mbox{else}~.
\end{aligned}
\end{equation}
Nevertheless, $\om =\sqrt{k^2+m^2} \in{\mathbb R}$, so that only
oscillating and no growing modes are excited.

Physically this result is not surprising. The distribution function of
relativistic particles cannot change faster than with the speed of
light and hence modes with frequencies $\om >k$ cannot be
excited. Therefore, only the tachyonic modes are relevant. However,
since $\Pi$ is not unstable, there is no instability. Only modes with
$k^2\geqslant -m^2$, and hence $\om\in\mathbb{R}$ are excited. We
expect this to hold true whenever the temporal change in the
perturbations is due to the motion of particles.

For vector and scalar perturbations, one obtains similar results. For
\textit{vector} perturbations, we have
\begin{equation}
\Pi_{\si\bullet}^{(V)} = \left\{\begin{array}{ll}
 \frac{-i}{2}\MM_{\si\bullet}(\om/k,k)\left(1-\frac{\om^2}{k^2}\right)
 \frac{\om}{k}
  & \mbox{ for }  \om^2<k^2~, \\
0 & \mbox{ else}~.  \end{array} \right.
\end{equation}
and for \textit{scalar} perturbations
\begin{equation}
\Pi^{(S)} = \left\{\begin{array}{ll}
 \frac{1}{4}\MM^{(S)}(\om/k,k)\left(\frac{1}{3}-\frac{\om^2}{k^2}\right)
  & \mbox{ for }  \om^2<k^2~, \\
0 & \mbox{ else}~.  \end{array} \right.
\end{equation}

The detailed junction conditions are somewhat involved, but they
evidently imply again that only the modes with $0\leqslant \om^2<k^2$
are excited,
\begin{equation}
 C_{\si\bullet} \neq 0 \quad \mbox{only if}\quad -k^2<m^2\leqslant 0~.
\end{equation}
Hence only tachyonic modes, $m^2<0$ with $\om\in {\mathbb R}$ are
present. But, since the distribution function $\MM$ does not grow
exponentially in time, truly unstable modes with $\om^2 =k^2+m^2<0$
are not allowed.

This example is interesting in the sense that the new tachyonic modes
play an important role, but no instability develops.

\subsection{Comments about cosmology}

In a cosmological setup, we assume the brane to move in \ads. The
scale factor is then given by $a=L/\ybr(\eta)$, where $\eta$ is
conformal time on the brane. The cosmological equations for the
unperturbed case are well known (see \EG \cite{Ringeval:2003na}),
\begin{align}
  \left(\frac{\dot a}{a}\right)^2 \equiv \HH^2 &=
   \frac{\ka_4}{3}\rho a^2\left(1+\frac{\rho}{2\tension}\right) ~,\\
\dot\rho +3\HH(\rho + P) &= 0 ~,
\end{align}
with $\ka_5\tension = 6/L$ and $\ka_4=\ka_5/L$. At low energy,
$\rho\ll \tension$ or $\HH \ll a/L$, we recover the standard Friedmann
equations.

Brane motion significantly alters the junction condition. For
instance, the perturbed junction condition for tensor
perturbations~\eqref{eq:2ndten} becomes
\begin{equation}\label{eq:Tjunc}
 L\HH\dd_tH_{\si\bullet}
 -a\left[1+ \left(\frac{L\HH}{a}\right)^2
 \right]^{1/2}\!\dd_yH_{\si\bullet}
 = \frac{\ka_5}{2}a^2\Pi_{\si\bullet}~,
\end{equation}
where $t$ and $y$ denote bulk coordinates. Clearly,
$H_{\si\bullet}(t,y)$ is no longer separable and we have to rely on
numerical simulations for its determination
(see~\cite{Koyama:2004cf,Hiramatsu:2004aa} for recent works on the
homogeneous case, $\Pi_{\si\bullet}\equiv 0$).

In a cosmology dominated by relativistic collisionless particles,
metric perturbations enter in the first order perturbation
Eq.~\eqref{eq:liouT2}, since $P=\rho/3$ is now a background
quantity. Hence $\Pi_{\si\bullet}$ and $H_{\si\bullet}$ must be
determined simultaneously via Eqs.~\eqref{eq:liouT2}, \eqref{eq:PiT}
and \eqref{eq:Tjunc}. It is not yet clear to us whether an instability
may develop in the presence of anisotropic stresses. A detailed study
is in preparation~\cite{CDR}.

\section{A simple orbifold model}
\label{RSmod}
\subsection{The toy model}
In Sec.~II we have found exponentially growing perturbations in the
scalar, vector and tensor sectors of a RSII background. These can be
generated from small, everywhere regular initial data. To complete the
discussion, we have to specify a brane equation of motion for the
anisotropic stresses and solve the full system. In the previous
section we have shown that in the RSII model linear perturbations are
still stable, but in a cosmological context, the coupling to metric
perturbations might induce an instability. Here we present, instead, a
simple toy model, to show that nonlinearities can lead to
instabilities already if the brane is flat Minkowski space.

We want to show the following: The instability which we have found is
not due to an instability in the equation of motion of the anisotropic
stresses on the brane nor to an instability of the bulk in absence of
a brane. It is also not coming from the choice of the wrong boundary
conditions (incoming wave, advanced instead of retarded solution,
\ETC), but it is due entirely to the singular orbifold construction
used in RSII.

Only the fact that we have two copies of $y\geqslant\ybr$ renders the
$K$ mode normalizable, which would otherwise diverge exponentially for
$y \ra 0$. In the same way one obtains an instability if one glues
together twice the same half of a simple Minkowski space. To see this
we consider four-dimensional Minkowski space-time, with orbifoldlike
spatial sections which can be identified with two copies of
$z\geqslant 0$. The $2+1$ dimensional ``brane'' is represented by the
plane $z=0$ and the ``bulk'' by two copies of $z>0$.

We consider a bulk field $\phi(t,\bxp,z)$, which satisfies the
ordinary hyperbolic wave equation,
\begin{equation}\label{eq:wave}
\Box\phi = -(\dd_t^2- \lap_{\parallel}- \dd_z^2)\phi=0~.
\end{equation}
Here $\bxp = (x,y)$ are the coordinates parallel to the ``brane'' and
$\lap_{\parallel}= \dd_x^2 +\dd_y^2$.  Like for the metric in RSII, we
require $\phi$ to be continuous across the brane ($z=0)$, but it may
have a kink, \IE its derivative may jump.

Confined on the brane is a field $\Pi$ which is given by
\begin{equation}\label{eq:Pijunction}
\lim_{z\ra 0_+}\dd_z\phi = \Pi~.
\end{equation}

Finally, we specify the equation of motion for $\Pi$ on the brane (the
matter equation):
\begin{equation}\label{eq:Pi}
(\dd_t^2 -\lap_{\parallel})\Pi =
-\frac{\dd_\mu\phi\dd^\mu\Pi}{\phi}~,
\end{equation}
where $(\dd_\mu) = (\dd_t,\dd_x,\dd_y)$ is the gradient on the brane.

This resembles a ``covariant derivative'' with $\dd_\mu\phi/\phi$
playing the role of the Christoffel symbols.

\subsection{Stable and unstable modes}
We first show that, if we do not allow for any ``singularity'', \IE
all fields and their first and second derivatives have to be
continuous everywhere, this system has no instability.

In this case, the fundamental solutions are simply
\begin{equation}
 \label{eq:nosing.phi}
 \phi = \phi_0\exp\left[\pm i(\om t-\bp\cd\bx)\right]~, \quad
 \om^2 - \bp^2 =0~,
\end{equation}
with $\bx = (x,y,z)= (\bxp,z) $ and $\bp = (p_x,p_y,p_z) =
(\bpp,\pp)$.  From Eq.~\eqref{eq:Pijunction} we obtain
\begin{equation}
 \label{eq:nosing.pi}
 \Pi = \mp i\pp \phi_0\exp\left[\pm i(\om t-\bpp\cd\bxp)\right]~.
\end{equation}
It is easy to verify that this is compatible with the equation of
motion~\eqref{eq:Pi} for $\Pi$. Clearly, these solutions do not
exhibit any instability. They do not grow, but oscillate in time.  For
an observer confined to the brane, $\phi$ appears as a field with mass
$m^2 = \pp^2\geqslant 0$. For $\pp^2= m^2$ fixed, this system can be
seen as a system of two fields $\Pi,~\phi_m$ on the brane, satisfying
\begin{equation}
 \label{eq:phi.brane}
 (\dd_t^2 -\lap_{\parallel})\phi_m = m^2\phi_m~,
\end{equation}
and Eq.~\eqref{eq:Pi}. This corresponds to a stable system of
scalar fields on the brane.

Now we proceed to the orbifold construction. We require $\Ze_2$
symmetry, $\phi(-z)=\phi(z)$, but allow $\dd_z\phi$ to jump at
$z=0$. All the equations of motion remain the same. Now, the previous
solutions for $\phi$ have to be combined into $\Ze_2$-symmetric linear
combinations,
\begin{equation}
 \label{eq:phiZ2}
 \phi = \phi_0\exp\left[\pm i(\om t-\bp\cd\bx)\right] +
        \phi_0\exp\left[\pm i(\om t-\bar{\bp}\cd\bx)\right]~,
\end{equation}
where $\bar{\bp} = (\bpp,-\pp)$ if $\bp = (\bpp,\pp)$. Such solutions
are still regular (continuous derivatives) at $z=0$.  However, there
now appears a new set of solutions:
\begin{align}
 \label{eq:phi.inst}
 \phi &= \phi_0\exp\left[\om t-i\bpp\cd\bxp -k|z|\right]~, \\
 \Pi  &= -k\phi_0\exp\left[\om t-i\bpp\cd\bxp\right]~,
 \label{eq:pi.inst}  \\
 \mbox{with}\quad \om^2 &=k^2 - \bpp^2~, \quad k>0~.
\end{align}
For $k^2>\bpp^2$, we have $\om^2>0$ and can thus choose $\om>0$, so
that these solutions grow exponentially in time. Evidently, $\phi$
obeys the bulk wave Eq.~\eqref{eq:wave} and $\Pi$
satisfies its equation of motion~\eqref{eq:Pi}.

From the brane point of view, $\phi$ is now a tachyonic field with
$m^2 = -\om^2-\bpp^2<0$. In the pure brane model (without a bulk) this
solution would therefore not appear. It is clearly only due to the
orbifold construction. Since $\phi$ decays exponentially away from the
brane, this solution also does not describe an ingoing wave.

\subsection{Green's function approach}
We finally want to address the question whether this solution may
contain ``a-causal'' contributions; or, more formally, whether it can
be constructed from the retarded Green's function alone.

We choose as Green's function for $\phi$ the one with
\begin{align}
 \xx&=(t,x,y,z)~,  \nonumber \\
 \Box G(\xx; \xx') &= \de^{(4)}(\xx-\xx')~, \label{eq:Green1}  \\
 \dd_{z'}G(\xx,\xx')\mid_{z'=0} &= 0~.   \label{eq:Green.bound}
\end{align}
Then, Green's formula gives (see, \EG \cite{Courant:1937})
\begin{equation}
 \phi(\xx) =
 \int_{z'=0}dt'd\bxp'\left[\dd_{z'}G(\xx,\xx')\phi(\xx')
   - G(\xx,\xx')\dd_{z'}\phi(\xx') \right]~.
\end{equation}
Using the boundary conditions at $z=0$ for $G$ and $\phi$,
this yields
\begin{equation}
 \label{eq:Green2}
 \phi(\xx) = -\int_{z'=0}dt'd\bxp'G(\xx,\xx')\Pi(t',\bxp')~.
\end{equation}
The analogous expression for the RSII model is given
in~\cite{Sasaki:1999mi}.

We now construct the retarded Green's function obeying the
boundary condition~\eqref{eq:Green.bound}. We perform a Fourier
transform in $t'$ and $\bxp'$ so that~\eqref{eq:Green1} becomes
\begin{equation}
 \left(\om^2-\bpp^2 +\dd^2_{z'}\right)\tilde G(z;z') = \de(z-z')~.
\end{equation}
The boundary condition~\eqref{eq:Green.bound} is satisfied by
the modes
\begin{equation}
 u_q(z) = \frac{1}{\sqrt{2}}\left(e^{iqz} + e^{-iqz}\right)~, \quad
 q^2 = \om^2-\bpp^2~.
\end{equation}
With the correct normalization (see, \EG \cite{Sasaki:1999mi}), we
obtain the retarded Green's function
\begin{align}
 \left.G(\xx,\xx')\right|_{z'=0}
 &= \frac{1}{\pi}\int_0^\infty dq
     \left(e^{iqz} + e^{-iqz} \right)
     \int\frac{d\om d^2\bpp}{(2\pi)^3}
     \nonumber \\
 &  \quad\times
     \frac{e^{-i\om(t-t')+i\bpp\cdot(\bxp-\bxp')}}
     {q^2+\bpp^2 -(\om+i\ep)^2}\nonumber\\
 &= 2\int\frac{d^4p}{(2\pi)^4}
     \frac{e^{ip_\mu(x^\mu-x^{\prime\mu})}}{\bp^2
     -(\om+i\ep)^2 } \nonumber \\
 &= \theta(t-t')\de\left(t\!-\! t'\!-\!\sqrt{z^2
     +|\bxp \!-\!\bxp'|^2}\right)  \nonumber \\
 &  \quad\times\frac{1}{(2\pi)\sqrt{z^2+|\bxp-\bxp'|^2}}~,
     \label{eq:Green5}
\end{align}
where $\theta$ is the Heaviside function,
\begin{equation*}
\theta (x) = \left\{\begin{array}{ll}
 1 & \mbox{ if } x\geqslant 0~, \\
 0 & \mbox{ if } x < 0 ~.\\
\end{array} \right.
\end{equation*}
Up to a factor $2$ which is due to $\Ze_2$ symmetry, this is just the
standard retarded Green's function of the four-dimensional wave
equation.

In order to show that our unstable solution~\eqref{eq:phi.inst} does
not invoke any a-causalities, \IE contributions from an advanced
Green's function, we now show that it can be obtained from the
retarded Green's function~\eqref{eq:Green5} by means of
Eq.~\eqref{eq:Green2}. We perform the calculations for $z>0$ and then
invoke $\Ze_2$ symmetry for $z<0$. Equations~\eqref{eq:pi.inst}
and~\eqref{eq:Green2} imply
\begin{align}
 \phi(t,\bxp,z)
 &= k\phi_0\int \frac{dt'd^2\bxp'}{2\pi}
     \frac{e^{\om t'-i\bpp\cdot\bxp'}}{\sqrt{z^2+\br^2}}\theta(t-t')
     \nonumber\\
 &  \quad\times\de\left(t-t'-\sqrt{z^2+\br^2}\right)~,
 \end{align}
where $\br = \bxp-\bxp'$.  With $d^2\bxp=d^2\br=rdrd\varphi$,
integration over $t'$ and $\varphi$ finally gives
\begin{align}
 \phi(t,\bxp,z)
 &= k\phi_0e^{\om t-i\bpp\cdot\bxp}\int_0^\infty
   \frac{dr\,re^{-\om\sqrt{z^2+r^2}}}{\sqrt{z^2+r^2}}
   J_0(|\bpp| r) \nonumber \\
 &= \phi_0e^{\om t-i\bpp\cdot\bxp -kz}~.
   \label{e:sol-phi}
\end{align}
Here $J_0$ is the Bessel function of order $0$. For the last equal
sign we used the integral No.~6.645.2 in~\cite{Gradshteyn:1965aa}.
This proves that $\phi$ from Eq.~\eqref{eq:phi.inst} is a purely
retarded solution and hence represents a true physical instability
of the system.

\section{Conclusions}
\label{sec:conclude}
We have shown that the perturbation equations of the RSII model allow
for tachyonic modes which can be exponentially unstable. We have then
argued that, within first order perturbation theory, the RSII model
remains stable. Nevertheless, we have given an example where the new
tachyonic modes play a crucial role. In the cosmological context, the
situation is significantly different and tachyonic modes could in
principle lead to instabilities. Since metric perturbations are no
longer separable, we have to rely on numerical simulations. We do not
yet know whether an instability may develop in the presence of
anisotropic stresses.  A more detailed study will be presented
elsewhere~\cite{CDR}.

Within a toy model, we have shown that the exponentially growing bulk
modes do not come from an instability of the brane equations of motion
or from the use of an inadequate Green's function, but are due to the
singular orbifold construction. As we show explicitly in our toy
model, the unstable modes can be obtained using the retarded Green's
function.  The toy model also shows that, taking into account
nonlinearities, even a flat Minkowski-space brane can become unstable.

Starting with small regular initial data on some hypersurface of
constant time, an exponential instability can build up in an \ads
orbifold. It remains to be examined whether this instability stays
exponential also in the cosmological context of a brane moving through
\ads, or if it disappears in an expanding universe like the Jeans
instability of Newtonian gravity.  This question is of uttermost
importance. If exponentially unstable modes are generated either on
the first or second order, this will inhibit the realization of
cosmology in terms of a RSII braneworld.

One may then go even further and ask whether such orbifold
constructions may not lead to instabilities in a much broader sense
than what has been discussed here.

\section*{Acknowledgment}
It is a pleasure to thank Roy Maartens, Ewald Roessl, Marcus Ruser, Misao
Sasaki, Misha Shaposhnikov and David Wands for helpful and
stimulating discussions. This work is supported by the Swiss
National Science Foundation and by the Fondation Birkigt.
 

\end{document}